\documentclass{article}
\usepackage{amsmath,amssymb,enumerate,eucal,pdfpages,theorem,xcolor}

\pdfoutput=1
\numberwithin{equation}{section}

\theoremheaderfont{\normalfont\scshape}
{\theorembodyfont{\rmfamily}}
{\theorembodyfont{\rmfamily}}
{\theorembodyfont{\rmfamily}}

\title{Hyperverse, 5-dimensional gravity and multiverses as nested Gogberashvili shells}
\author{Igor Yu. Potemine \\ 
\small Institut de Mathématiques, Universit\'e Paul Sabatier, Toulouse, France}
\date{}

\begin{document}
\maketitle
\abstract{We consider the \emph{Hyperverse} as a collection of multiverses in 5-dimen\-sional spacetime with gravitational constant $G$. Each multiverse in our simplified model is a bouquet of nested spherical Gogberashvili shells. If $g_k$ is the gravitational constant of a thin shell $S_k$ and $\varepsilon_k^{}$ its thickness then $G\sim 
\varepsilon_k^{}g_k^{}$. The physical universe is supposed to be one of those shells inside the local nested bouquet called \emph{Local Multiverse}. We relate this construction to Robinson-Trautman metrics describing expanding spacetimes with spherical gravitational waves.
Supermassive astronomical black holes, located at cores of elliptic/spiral galaxies, are also conjecturally described within this theory. Our constructions are equally consistent with the modern theory of cosmological coupling.}

\medskip

\tableofcontents

\section{Introduction}

According to the \emph{Newton's shell theorem}, the gravitational force, exerted on any object inside a hollow spherical shell, is zero. Consequently, in the case of nested hollow shells, one can ignore all spherical shells of greater radius.

\smallskip One century ago Kaluza and Klein constructed a 5-dimensional gravitation theory unified with electromagnetism. However, the extra fifth dimension in this theory is curled up to an unobservable scale.

\smallskip An alternative model with a true higher-dimensional Hyperverse where the matter is trapped on a 4-dimensional domain wall ($D$-brane), appeared, \emph{e.g.}, in the paper by Rubakov and Shaposhnikov \cite{RSh}. 

\smallskip Gogberashvili constructed an exact Schwarzschild-like solution of 5-dimensional Einstein equations, exhibiting an expanding spherical shell \cite{Gog}. He also solved the \emph{hierarchy problem}, reducing the particle theory to the same scale $G$ (5-dimensional gravity constant). 

\smallskip In addition, the trapping of matter on this shell is gravitationally repulsive and the expansion of Gogberashvili's spherical shell is accelerating, solving also the problem of dark energy.

\smallskip In my recent paper \cite{Pot}, it was argued that our Local Multiverse $\mathcal{M}$ can be considered as a time-amalgamated product 
of spacetimes. Here, first of all, we pursue this idea representing the 4-dimensional stratum of $\mathcal{M}$ as a bouquet of nested Gogberashvili shells.

\smallskip Cosmological constants around subsequent shells satisfy some recursive equations reflecting a certain \emph{harmony of the spheres}
(\emph{cf.}~sect 4).

\smallskip In the second part of this article we discuss intriguing relationships between astronomical black holes, spherical gravitational waves, multi-solitons/gyratons and Gogberashvili multiverses.

\section{5-dimensional Einstein equations}

Consider the 5-dimensional Einstein equation in the following form:
\begin{equation}
R_{\mu\nu}-\frac{1}{2}g_{\mu\nu}R = \Lambda g_{\mu\nu}+6\pi^2GR_{\mu\nu}
\end{equation}
Gogberashvili searched for a Schwarzschild-like solution of the form:
\begin{equation}
ds^2=A(r)dt^2-A(r)^{-1}dr^2-r^2d\Omega^2,
\end{equation}
where $r$ is the 4-dimensional radial coordinate and $d\Omega^2$ is the
3-dimensio\-nal volume element \cite[eq.~(2)-(4)]{Gog}.

\smallskip The solution is given in terms of metrics of inner and outer regions:
\begin{align}
ds_{+}^2 &=(1-2MG/r^2+\Lambda_{+}r^2)dt^2-\\
&-(1-2MG/r^2+\Lambda_{+}r^2)^{-1}dr^2-r^2d\Omega^2,\\
ds_{-}^2 &=(1+\Lambda_{-}r^2)dt^2-(1+\Lambda_{-}r^2)^{-1}dr^2-
r^2d\Omega^2,
\end{align}
separated by a time-like 4-dimensional spherical shell (brane, bubble) with FLRW metrics:
\begin{equation}
ds^2 = d\tau^2-a^2(\tau)d\Omega^2,
\end{equation}
where $\tau$ is the intrinsic time of this spherical universe 
\cite[eq.~(13)-(14)]{Gog}.

\section{Nested Gogberashvili models}

It is natural to consider the nested variant of considered solution as a simplified model of \emph{multiverses} restricted to their $(3+1)$-dimensional strata. So, we consider a sequence of $n$ nested spherical shells-universes $S_k$ around the same center with masses $m_k^{}$, $1\leqslant k\leqslant n$. Denote by $\displaystyle M_k=\sum_{i=1}^k 
m_k^{}$ the total mass of first $k$ shells. Local outer and inner metrics are defined by the following formulas:
\begin{align}
ds_{+,k}^2 &=(1-2M_kG/r^2+\Lambda_{+,k}r^2)dt^2-\\
&-(1-2M_kG/r^2+\Lambda_{+,k}r^2)^{-1}dr^2-r^2d\Omega^2,\\
ds_{-,k}^2 &=(1-2M_{k-1}G/r^2+\Lambda_{+,k-1}r^2)dt^2-\\
&-(1-2M_{k-1}G/r^2+\Lambda_{+,k-1}r^2)^{-1}dr^2-r^2d\Omega^2.
\end{align}
Thus, $ds_{-,k}^2=ds_{+,k-1}^2$ for $2\leqslant k\leqslant n$ . We put here $M_0=0$, so the innermost Gogberashvili metric has the form:
\begin{equation}
ds_{-,1}^2 =(1+\Lambda_{-}r^2)dt^2-(1+\Lambda_{-}r^2)^{-1}dr^2-r^2d\Omega^2,
\end{equation}
where, by definition, $\Lambda_{-}=\Lambda_{+,0}$.

\section{Harmony of transcosmic spheres}

Let $g_k$ denote the gravitational constant on the $k$th shell of thickness  $\varepsilon_k^{}$ then $G\sim\varepsilon_k^{}g_k^{}$. It is natural to suppose the following relations:
\begin{equation}
\Lambda_{+,k-1} - \Lambda_{+,k} = \sigma_k^{} g_k^{}/G,
\end{equation}
$1\leqslant k\leqslant n$, where $\sigma_k^{}$
is the energy density of the $k$th shell \cite[eq.~(21)]{Gog}. Consequently,
\begin{equation}
\Lambda_{+,n} = \Lambda_{+,0} - \sum_{k=1}^{n}\sigma_k^{} g_k^{}/G.
\end{equation}
In this picture, a large 5-dimensional gravitational constant 
$\Lambda_{+,0}$ could generate a sequence of potential wells
(cf.~\cite[fig.~1]{RSh}), trapping matter on a sequence of expanding branes-universes. The expansion rate of the bubble $S_k$ depends on its energy density as well as on $\Lambda_{+,k-1}$ and $\Lambda_{+,k}$. Meanwhile, the 4-dimensional cosmological constant on $S_k$ vanishes.

\includepdf[pages=1, scale=.5]{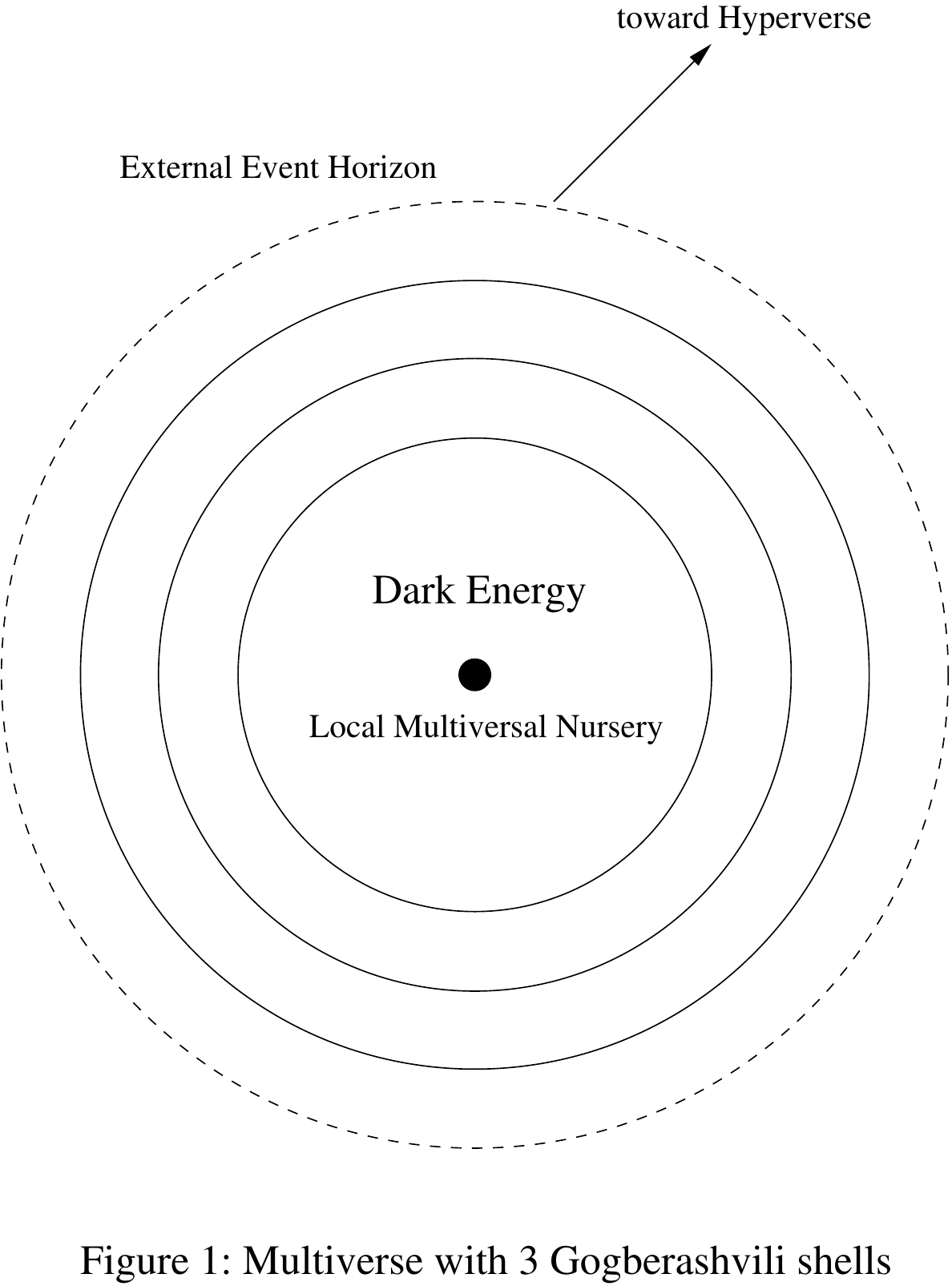}

\section{Sequence of event horizons}

The metric $ds_{+,k}^2$ becomes singular when
\begin{equation}
1-2M_kG/r^2+\Lambda_{+,k}r^2 = 0
\end{equation}
So, we have the following growing sequence of event horizons:
\begin{equation}
r_k^{}=\sqrt{\left(\sqrt{1+8GM_k\Lambda_{+,k}}-1\right)/2\Lambda_{+,k}}
\end{equation}
for $1\leqslant k\leqslant n$.

\smallskip The most meaningful is the \emph{external event horizon} with radial coordinate $r_n$, located outside of $S_n$.

\section{Time-amalgamated multiverses}

Four-dimensional Gogberashvili shells $S_k$ with FLRW metrics
\begin{equation}
ds_k^2 = d\tau_k^2-a^2(\tau_k^{})d\Omega^2,
\end{equation}
$1\leqslant k\leqslant n$, can be glued into a multiverse $\mathcal{M}$ according to the procedure of time amalgamation, described in my paper \cite{Pot}. It means that intrinsic times $\tau_k^{}$ become mutually synchronized and there is a common timeline $\mathbb{T}$ such that
\begin{equation}
\mathcal{M}=S_1\times_{\mathbb{T}}^{}\times\cdots\times_{\mathbb{T}}^{}S_n
\end{equation}
Here $\mathcal{M}$ is considered as a fibered product.

\smallskip In this way, all shells $S_k$ can be studied simultaneously.

\section{Dark energy and multiversal nurseries}

It is possible to endow the Hyperverse with a metric approximately describing several widely separated Gogberashvili multiverses. For example, one can use the work of Kashif Alvi on widely separated binary black holes
\cite{Alv}.

\smallskip In our picture, the Hyperverse has numerous powerful sources of dark energy with large local cosmological constants. Those sources play role of the \emph{nurseries of multiverses}.

\smallskip The situation is somewhat similar to the star formation and stellar nurseries in galaxies. In the latter case, it is related to the collapsing matter in giant molecular clouds.

\smallskip Notice that, in our constructions, the dark energy is a higher-dimensio\-nal phenomenon of the Hyperverse.

\section{Black holes as multiverses}

For an external hyperversal observer, any $(3+1)$-dimensional 
Gogberashvi\-li multiverse would eventually appear as a supermassive black hole. It suggests the idea that astronomical black holes in our physical universe are, actually, nested and spherically embedded $(2+1)$-dimensional branes-universes.

\smallskip All solutions to $(2+1)$-dimensional Einstein equations were actually classified \cite{PSM}. Gogberashvili models correspond in this case to spherical Robinson-Trautman solutions \cite{RT}. In canonical coordinates the Robinson-Trautman metric is given by the formula:
\begin{equation}
ds^2 = \frac{r^2}{P^2}dx^2+2(er^2+f)dudx-2dudr+ F(a,e,P,\Lambda;r)du^2,
\end{equation}
where $a(x,u)$, $e(x,u)$, $f(x,u)$ and $P(x,u)$ are some metric functions while $F(a,e,P,\Lambda;r)$ is a polynomial of degree 2 with respect to $r$ depending on $a$, $e$, $P$ and the cosmological constant $\Lambda$
\cite[eq.~(37)]{PSM}. 

\smallskip Similar solutions in the form
\begin{equation}
ds^2 = g_{pq}dx^pdx^q+2g_{up}dudx^p-2dudr+g_{uu}du^2,
\end{equation}
exist in all $(d+1)$-dimensional spacetimes with $d\geqslant 2$ 
(\emph{cf}.~\cite[eq.~(1)]{PS}).

\smallskip We should also mention innovative works of T.X. Zhang on black hole universes (\emph{cf.}~\cite{Zha} and references therein).

\section{Multi-solitonic and gyratonic gravitational waves}

So, we are interested in Robinson-Trautman metrics describing expanding spacetimes with spherical gravitational waves. Multi-soliton solutions
of this type, associated with matter trapping, correspond to our Gogberashvili multiverses.

\smallskip If we admit that the null matter field has some internal spin / angular momentum, there are also more general (multi-)gyratonic solutions. They give rise to rotating Gogberashvili models/multiverses.

\smallskip This intriguing topic, including a thorough comparison ot two models, should be studied in more details. In particular, the harmony of transcosmic spheres should acquire more complete meaning.

\section{Elliptic and spiral galaxies}

It seems that it partially explains the classification of galaxies, harboring supermassive black holes at their cores. Indeed, on the one hand, supermassive black holes of multi-solitonic type could generate elliptic galaxies. 

\smallskip On the other hand, multi-gyratonic black holes could generate spiral galaxies. The number of spirals and their configuration is related in this case to the number of multiversal layers and the harmony of the spheres (\emph{cf.}~sections 4 and 9).

\smallskip Relativistic jets, emitting by supermassive black holes at galaxy cores, also suggest their gyratonic nature.

\section{Cosmological coupling}

Our constructions are consistent with the modern theory of \emph{cosmological coupling} \cite{Bis}.

\smallskip Recent astronomical observations show the abundance of black holes with unexpectedly high masses. Croker \emph{et al.} proposed a
mechanism, explaining the growth of black holes through anomalous coupling of matter with gravity \cite{Cro}.

\smallskip It implies that the expansion of the universe and a large (but hidden) cosmological constant contribute to the growth of black hole masses.

\section{Conclusion}

It turns out that nested Gogberashvili shells represent good models for multiverses inside the Hyperverse.

\smallskip In this paper we have restrained the Hyperverse to its local $(4+1)$-dimensional stratum. However, similar solutions exist in all dimensions $d+1$ with $d\geqslant 2$. In fact, we can construct an infinite tower of embedded Gogberashvili multiverses of various dimensions.

\smallskip An alternative approach is given by Robinson-Trautman metrics describing expanding spacetimes and spherical gravitational waves.

\smallskip It provides us with an intriguing idea to consider supermassive astronomical black holes as expanding $(2+1)$-dimensional multiverses (with possible higher strata too).

\smallskip It might give new angles of view on the black hole growth, the cosmological coupling, relativistic jets and even on the classification of galaxies.

\bibliographystyle{unsrt} 
\bibliography{hyperverse}

\bigskip
\begin{flushright}
Igor Potemine\\
Institut de Mathématiques\\
Universit\'e Paul Sabatier\\
118, route de Narbonne\\
31062 Toulouse (France)
\end{flushright}

\begin{flushright}
e-mail: igor.potemine@math.univ-toulouse.fr
\end{flushright}

\end{document}